\documentclass[fleqn,usenatbib,useAMS]{mnras}

\usepackage{graphicx}	
\usepackage{wrapfig}
\usepackage{lscape}
\usepackage{rotating}
\usepackage{amsmath}	
\usepackage{amssymb}	
\usepackage{multicol}        
\usepackage{bm}		
\usepackage{pdflscape}	
\usepackage{natbib}
\usepackage{hyperref}
\usepackage{multirow}
\usepackage{arydshln}
\usepackage{float}
\usepackage{subfigure}
\usepackage{lineno}

\newcommand{\kms}{\,km\,s$^{-1}$} 

\usepackage[T1]{fontenc}
\usepackage{ae,aecompl}

\title[]{A Spectrophotometric analysis and dust properties of classical nova V5584 Sgr}
 \author[M. S. Bisht et al.]{
 Mohit Singh Bisht$^{1}$,
 A. Raj$^{1,2}$\thanks{E-mail: ashishpink@gmail.com },
 F.M. Walter$^{3}$,
 D. Bisht$^{1}$,
 Gargi Shaw$^{4}$,
 K. Belwal$^{1}$,
 \newauthor
 S. Biswas$^{1}$
 \\
 $^{1}$Indian Centre for Space Physics, 466 Barakhola, Netai Nagar, Kolkata 700099, West Bengal, India\\
 $^{2}$Uttar Pradesh State Institute of Forensic Science (UPSIFS), Aurawan, P.O. Banthra, Lucknow 226401 (U.P), India\\
 $^{3}$Department of Physics and Astronomy, Stony Brook University, Stony Brook, NY 11794-3800, USA\\
 $^{4}$Department of Astronomy and Astrophysics, Tata Institute of Fundamental Research, Mumbai 400005.,
India\\
 }

\date{Last updated XXX ; in original form YYY}
\pubyear{2024}

\begin{document}
\label{firstpage}
\pagerange{\pageref{firstpage}--\pageref{lastpage}}
\maketitle

\begin{abstract}
In this work, optical observations of the nova V5584 Sgr are presented. These observations cover different phases including pre-maximum, early decline, and nebular. The spectra are dominated by hydrogen Balmer, Fe II, and O I lines with P-Cygni profiles in the early phase, which are subsequently observed in complete emission. The presence of numerous Fe II lines and low ejecta velocity aligns with the Fe II type nova classification. From optical and NIR colors it is clear that this nova manifests dust formation in the ejecta. The dust temperature and mass were estimated from a spectral energy distribution (SED) fit to the JHK band magnitudes and the WISE data.
Light curve analysis shows t$_2$ and t$_3$ values of $\sim$ 26 and $\sim$ 48 days, classifying the nova as moderately fast.
The physical and chemical properties during early decline and later phases were evaluated using the photoionization code \texttt{CLOUDY}. The best-fit model parameters from two epochs of multiwavelength spectra are compatible with a hot white dwarf source with a roughly constant luminosity of $\sim$ (2.08 $\pm$ 0.10) $\times$ 10$^{36}$ erg s$^{-1}$. We find an ejected mass of $\sim$
(1.59 $\pm$ 0.04) $\times$ 10$^{-4}$M$_{\odot}$. Abundance analysis indicates that the ejecta is significantly enriched relative to solar values, with O/H = 30.2, C/H = 10.8, He/H = 1.8,  Mg/H = 1.68, Na/H = 1.55, and N/H = 45.5 in the early decline phase, and O/H = 4.5, Ne/H = 1.5, and N/H = 24.5 in the nebular phase. 
\end{abstract}





\section{Introduction}
 
Novae are systems comprising interacting binaries. The primary star in these systems is a white dwarf (WD) that accretes hydrogen-rich material from a nearby companion star, known as the donor. If the donor is a main sequence or subgiant star, the WD accretes mass through Roche lobe overflow.
Alternatively, if the donor is a red giant, the mass transfer is driven by the stellar wind \citep{darnley2012progenitors,Darnley:2021ph}. The material lost from the companion star accumulates in an accretion disk around the WD before falling to its surface. 
This process forms a surface layer of hydrogen, helium, and other heavy elements on the WD. Due to intense gravity, the pressure at the base of
this layer increases and the material becomes electron degenerate.
When the temperature at the bottom of this degenerate layer reaches the Fermi temperature ($\sim$ 7 $\times$ 10$^{7}$ K), the layers expand rapidly, leading to a further increase in temperature that culminates in a thermonuclear runaway reaction (TNR) \citep{starrfield2016thermonuclear}. This reaction releases an enormous amount of energy, resulting in a nova outburst. 

The eruption results in a sudden rise in optical brightness, ranging from 8 to 18 magnitudes within 1–2 days, with peak luminosities reaching $10^4$ -- $10^5$ L$_\odot$, as suggested by
\citep{2012BASI...40..443J}.
Approximately 10$^{-4}$ to 10$^{-5}$ solar masses (M$_\odot$) of material is ejected into the interstellar
medium (ISM), typically at velocities ranging from a few hundred to several thousand kilometers per second \cite[][and references therein]{BodeEvansBook2008,Jose2020,Starrfield2020}. Classical Novae (CNe) significantly contribute to the Galactic abundances of $^{13}$C, $^{15}$N, and $^{17}$O and may also enrich other intermediate-mass isotopes, such as $^{7}$Be, $^{7}$Li and $^{26}$Al \citep{tajitsu2015explosive,izzo2015early,Starrfield2020}.

CNe are well known to form dust during outbursts, generally about 10-100 d after the eruption \citep{williams2013tdt2}. 
Observations have also provided evidence for the formation of molecules in the nova ejecta. As the ejected material cools and evolves, diatomic molecules such as CO \citep{rudy2003near,raj2012v496}, C$_2$ \citep{kawakita2017mid}, and CN \citep{wilson1935cyanogen} form. CNe containing Carbon-Oxygen White Dwarfs (CO novae) generally form dust within the ejecta. In these dust-forming novae, carbon often contributes a major component of the grains. In addition, other components such as silicates, polycyclic aromatic hydrocarbons (PAH), and silicon carbide (SiC) are frequently found \citep{gehrz1998nucleosynthesis,2012BASI...40..213E}.

Nova V5584 Sgr was discovered on October 26.439 UT, 2009, by Nishiyama and Kabashima, with a magnitude of 9.3 on two 60-sec unfiltered CCD frames \citep{2009CBET.1994....1N}. No object was visible at this location in their survey frames from October 20.449 UT, 2009 (limiting magnitude 13.9), and October 21.451 UT, 2009 (limiting magnitude 13.4). 
\cite{2009IAUC.9089....1coriil} did not reveal any object at the position of V5584 Sgr in their Palomar plate (limiting mag 21.0) on Oct. 26.764 UT, 2008.
\cite{2009CBET.1995_Kinugasa} reported early low-resolution optical spectroscopic observations on October 27.4 UT, 2009, showing H$\alpha$ emission with a full width at half maximum (FWHM) of about 600 \kms and a P-Cygni profile, indicating the object was an early-stage nova. The H$\alpha$ absorption minimum was blue-shifted by 900 \kms relative to the emission peak. P-Cygni profiles were also observed in the Fe II (multiplet 42) lines. \cite{2009CBET.1995_Mahera} obtained another low-resolution spectrum on October 27.42 UT, 2009, showing a similar H$\alpha$ profile, confirming the object as a classical nova. Further observations by \cite{2009CBET.1999_munari} on October 28.73 UT and October 29.72 UT, 2009, provided low, medium, and high resolution spectra. On 28.73 UT, the spectrum was dominated by a well-developed, highly reddened absorption continuum. The absorption lines indicated a FWHM of 310 \kms, and the heliocentric radial velocity was -283 \kms. The emission components were red-shifted by 440 \kms relative to the absorption spectrum. The high-resolution spectrum from October 29.72 UT showed a strong absorption continuum and very weak emissions from the Balmer and Fe II multiplets. These observations suggested that the nova is an Fe II type, approaching its optical peak.

\cite{2010russel_atel} reported dust formation within the nova ejecta based on near-infrared (NIR) observations taken on February 10, 2010 (106 days after discovery). The exact onset of dust formation was unknown due to the solar conjunction, but they estimated the dust temperature to be 880 $\pm$ 50 K. 
The NIR spectrophotometric observations from October 28.59 UT to November 8.56 UT, 2009, were conducted by \cite{raj2015near}. 
The early spectra displayed prominent P- Cygni profiles for H I, O I, C I, and N I lines. On November 5.64 UT, 2009 (10.2 days after discovery), they reported the first overtone CO bands in emission within the range of 2.29-2.40 \micron. The CO mass and the gas temperature (T$_{co}$) were estimated to be 3 $\times$ 10$^{-8}$ M$_\odot$ and 3500 K, respectively.

In this paper, we examine the evolution of the optical spectra of V5584 Sgr from October 28, 2009,
to October 17, 2010, utilizing data from the SMARTS database.
We used the photoionization code \texttt{\texttt{CLOUDY}} \citep{chatzikos2023} to analyze the observed spectra and develop a simple phenomenological model based on a spherical geometry to estimate crucial physical and chemical parameters of the nova. The paper is organized as follows. Section~\ref{observations} presents the details of the observations. Section~\ref{analysis} provides a detailed analysis of the spectrophotometric data and the estimation of key physical and chemical parameters using photoionization modeling and examining dust properties using WISE data. Sections~\ref{Discussion} and~\ref{summary} conclude with a comprehensive discussion and
summary, respectively.

\section{Observations}\label{observations}
Low-dispersion optical spectra and photometric data were acquired utilizing the Small and
Moderate Aperture Research Telescope System (SMARTS\footnote{\href{http://www.astro.sunysb.edu/fwalter/SMARTS/NovaAtlas/}{http://www.astro.sunysb.edu/fwalter/SMARTS/NovaAtlas/}}) facilities. Nine spectra were
obtained with the SMARTS R/C spectrograph between October 28, 2009, and October 17,
2010, under various sky conditions. The SMARTS R-C grating spectrograph, data reduction techniques, and observing modes are described by \cite{wal12}.
The log of spectroscopic observations is given in Table \ref{log}.

\begin{table}
	\caption{Observational log for spectroscopic data obtained for V5584 Sgr.}
	\label{log}
 
    {\begin{tabular}{cccccc}
			\hline
			\hline
			& \textbf{Time since}  & \textbf{Exposure}  &   \textbf{Wavelength}  &\textbf{Resolution}  \\
			\textbf{Date (UT)} & \textbf{discovery} & \textbf{time} & \textbf{range} & \textbf{(\AA)} \\
			& \textbf{(days)} & \textbf{(s)} &  \textbf{(\AA)} & \\
			\hline
			2009 Oct 28.02 & 1.58 & 600  & 6006--9482 & 8.7 \\[0.25ex]
			2009 Oct 28.99 & 2.55 & 900  & 6006--9491 & 8.7 \\[0.25ex]
			2009 Oct 29.99 & 3.56 & 300  & 6005--9497 & 8.7 \\[0.25ex]
			2009 Oct 31.28 & 4.60 & 800  & 5628--6949 & 3.1 \\[0.25ex]
			2009 Nov 02.01 & 6.57 & 600  &5631--6951 & 3.1 \\[0.25ex]
			2009 Nov 06.03 & 10.59 & 600  & 3652--5423 & 4.1 \\[0.25ex]
			2009 Nov 07.00 & 11.56 & 720  & 6226--8870 & 6.6 \\[0.25ex]
			2010 Aug 15.18 & 292.74 & 2700   & 2797--9610 & 17.0 \\[0.25ex]
			2010 Oct 17.02 & 355.58 & 2700   & 2705--9508 & 17.0 \\[0.25ex]
			\hline
	\end{tabular}}
\end{table}

\section{Analysis}\label{analysis}

\subsection{Optical and NIR light curve, reddening and distance}\label{reddening and distance}

\begin{figure}
  \includegraphics[width=1.0\columnwidth]{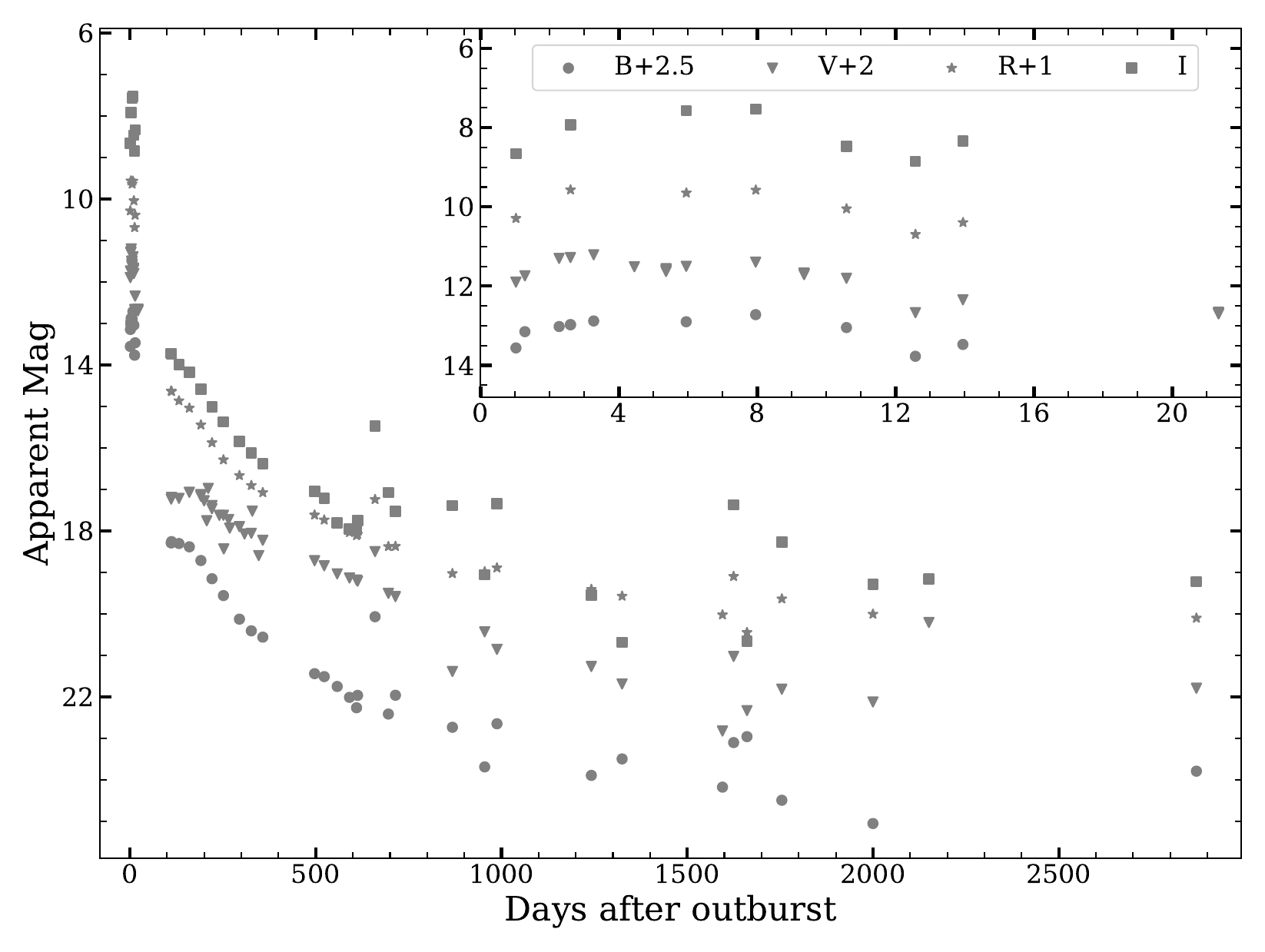}
 \caption{Optical light curves of V5584 Sgr generated using data taken from AAVSO, SMARTS, and reported magnitudes. Offsets have been applied for clarity to all the magnitudes except $I$.}
 \label{lc}
\end{figure}

The optical light curves based on the data from the American Association of Variable Star Observers (AAVSO\footnote{\url{https://www.aavso.org/}}) database and SMARTS are presented in Fig. \ref{lc}.

\begin{figure}
 \includegraphics[width=1\columnwidth]{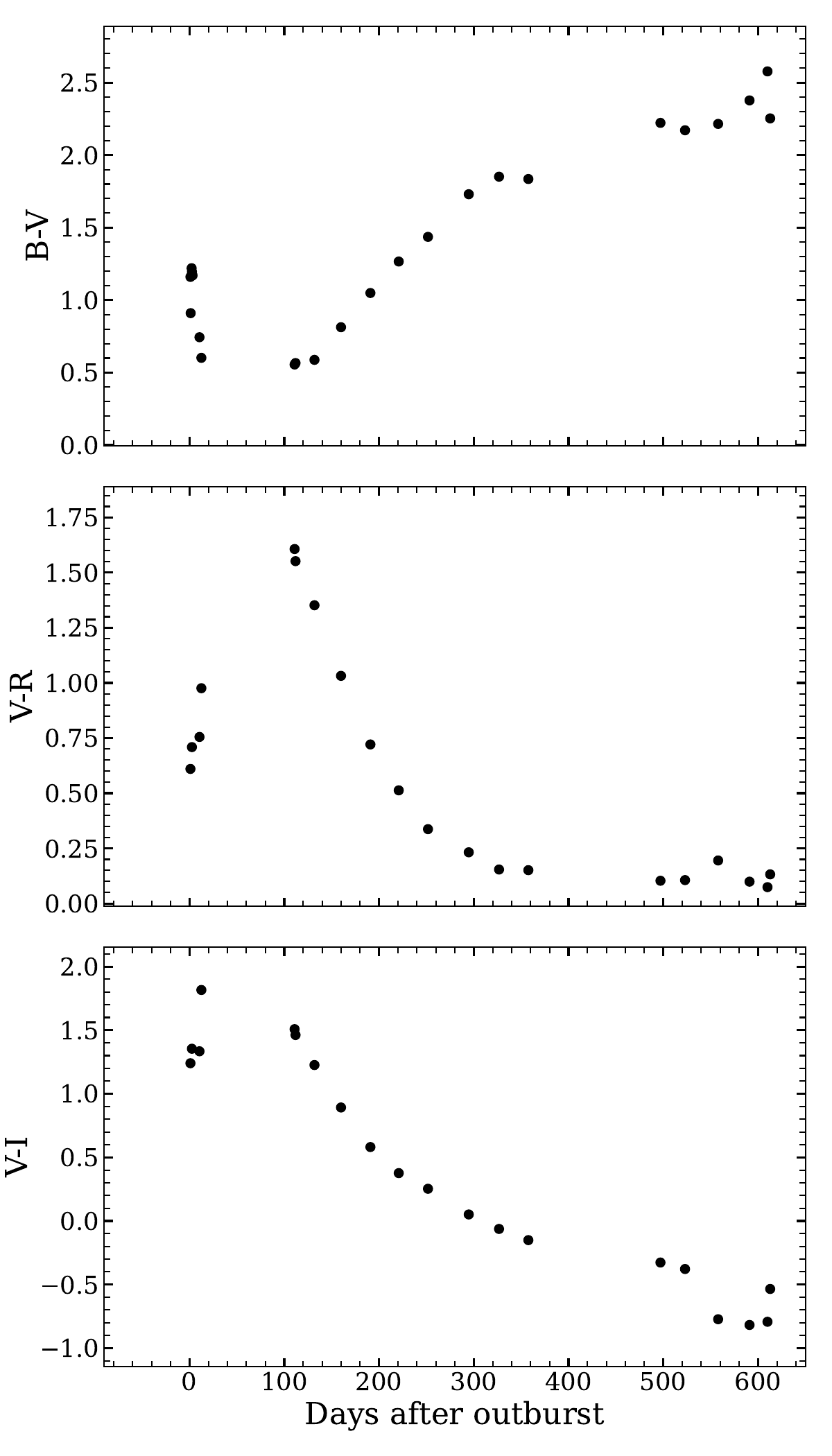}
 \caption{Evolution of optical color terms of nova V5584 Sgr from day 1 (pre-maximum phase) to 650 (decline phase) since discovery.}
 	\label{colorterm}
\end{figure}

We consider the day of discovery as the day of outburst (t$_0$). The BVRI band light curves start within a day of the discovery. The nova reached its maximum value of $V\rm_{max}$ = 9.21 mag on October 29.708 UT, 2009 (3.26 days) \citep{2009CBET.1999_munari}. 
The brightness in the BVRI bands exhibits a slow decline after reaching the peak. By applying a
least-squares regression fit to the $V$-band light curve, we determine t$_2$ to be 26 $\pm$ 1 days.
Due to the limited coverage of the photometric data resulting from the solar conjunction, we estimated t$_3$ to be 48 $\pm$ 2 days using the relation t$_3$ = 2.75 (t$_2$)$^{0.88}$ by \cite{1995cvs..book.....W}. Based on the t$_2$ value the nova belongs to a moderately fast nova category \citep{1957gano.book.....G}. 
We calculated the absolute magnitude of the nova to be M$_v$ = -7.60 $\pm$ 0.1 using the maximum magnitude versus rate of decline (MMRD) relation given by \cite{della1995calibration}. However, the MMRD relation may deviate for fast and faint novae \citep{kasliwal2011discovery,shara2017hubble}, introducing some uncertainty into our estimate. Using the above values, we estimate a distance modulus of Vmax - M$_v$ = 16.8 mag.
By utilizing the values of M$_v$ and t$_3$ and applying them to the equation 2 and 6 given in \cite{liv92}, we obtained the mass M$_{WD}$ = 0.92 $\pm$ 0.07M$_\odot$. This indicates that the nova V5584 Sgr contains a low-mass CO WD (M$_{WD}$ $\leq$ 1.2M$_\odot$ ). 
The outburst luminosity is calculated using the formula $M_{bol}$ = 4.8 + 2.5log (L/$L_{\odot}$), where the
bolometric correction applied to $M_V$ is assumed to range from -0.4 to 0.00, corresponding to
spectral types A to F, respectively. Novae at maximum typically exhibit a spectral type between A
and F \cite{gehrz1988infrared}.
Using $M_V$ = -7.6, we calculate the luminosity of the outburst (1.12 $\pm$ 0.20)$\times$ $10^5$ $L_{\odot}$. 
\cite{van1987ubv} derived a mean intrinsic color (B-V)$_o$ =+0.23 $\pm$ 0.06 for novae at maximum. 
Using this value, we calculated \textit{E(B-V)} from the observed (B-V) color of 1.18, resulting in \textit{E(B-V)} = 0.95 $\pm$ 0.06. 
The previously reported reddening values are: 0.94 \citep{raj2015near}, 0.82 \citep{poggiani2011spectroscopic}, and 0.75 \citep{hachisu2021ubv}. 
We adopted the average reddening value \textit{E(B-V)} = 0.80. 
Using the reddening value, we derived the interstellar extinction as A$_v$ = 2.48 for R = 3.1. Applying the above parameters to the distance modulus relation, we calculated the distance to be 7.4 $\pm$ 0.5 kpc. 

The color evolution B-V, V-R, and V-I is presented in Fig. \ref{colorterm}. On day 1, the B-V, V-R, and V-I colors were approximately 1.16, 0.61, and 1.24, respectively.
By day 110, the B-V color had decreased to 0.56, while V-R and V-I had increased to 1.60 and 1.51, respectively. After day 110, B-V started to increase, reaching 2.58 by day 609, while V-R and V-I decreased to 0.07 and -0.80, respectively. Initially, when the nova is optically thick, changes in the optical colors largely reflect changes in the extinction due to the formation and subsequent thinning of dust. However, once in the nebular phase line emission dominates over the optical continuum in the B, V, and R bands. The apparent reddening in the B-V color is largely attributable to the increasing brightness of the [O III] 4959/5007 lines, which dominate in the V band. The V-R color drops as H$\alpha$ fades relative to
[O III]. V-I becomes bluer largely because the continuum, which dominates in
the I band, fades faster than H$\alpha$. On the final observation (day 2870) of optical light curve, the nova had magnitudes of 21.3, 19.9, 19.1, and 19.2 in the B, V, R, and I bands, respectively.

\begin{figure}
  \includegraphics[width=1.0\columnwidth]{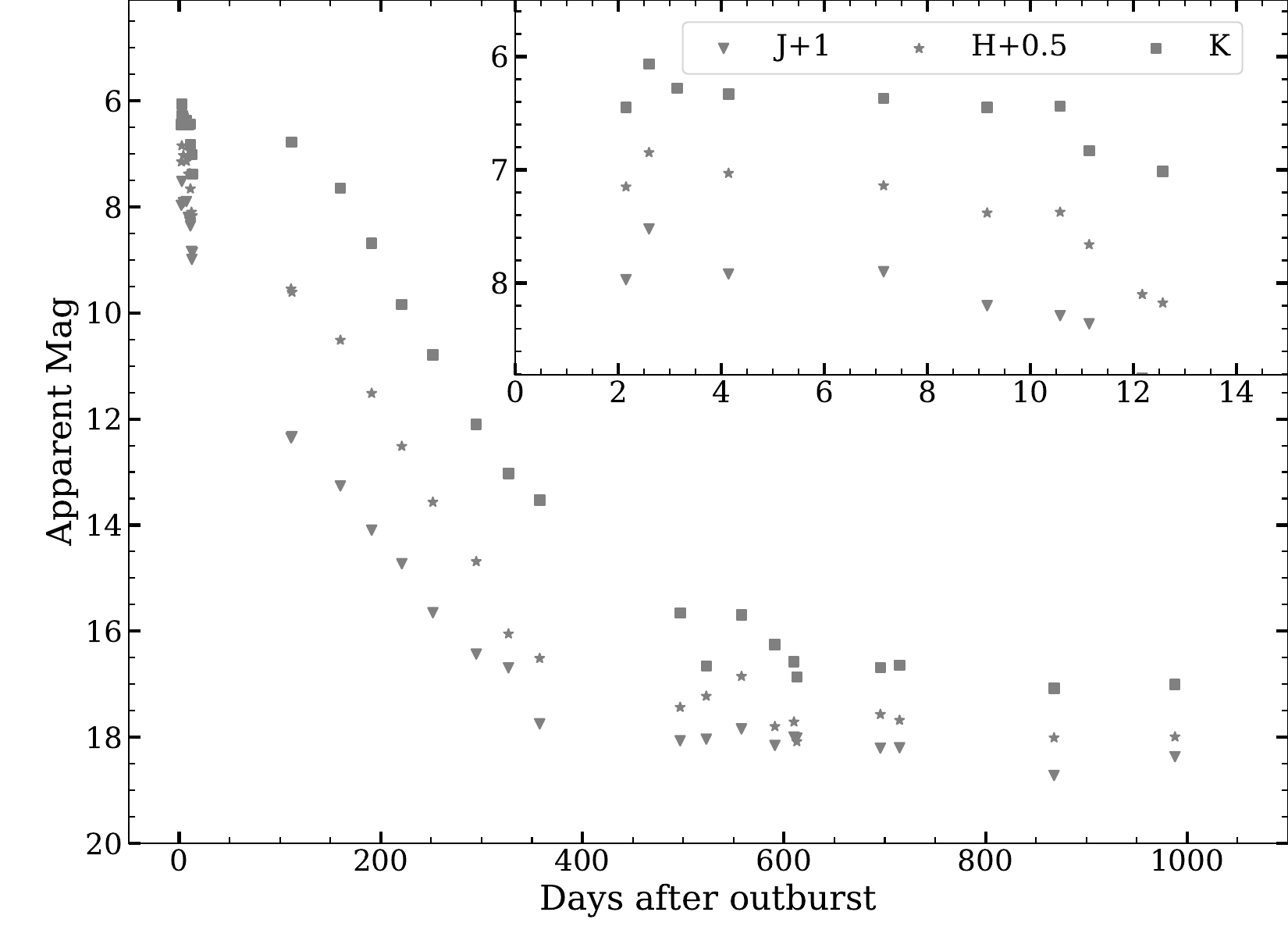}
  \caption{NIR light curves of V5584 Sgr generated using JHK band data from SMARTS and \citet{raj2015near}. Offsets have been applied for all the magnitudes except $K$ for clarity.}
  \label{nirlc}
\end{figure}

\begin{figure}
 \includegraphics[width=1\columnwidth]{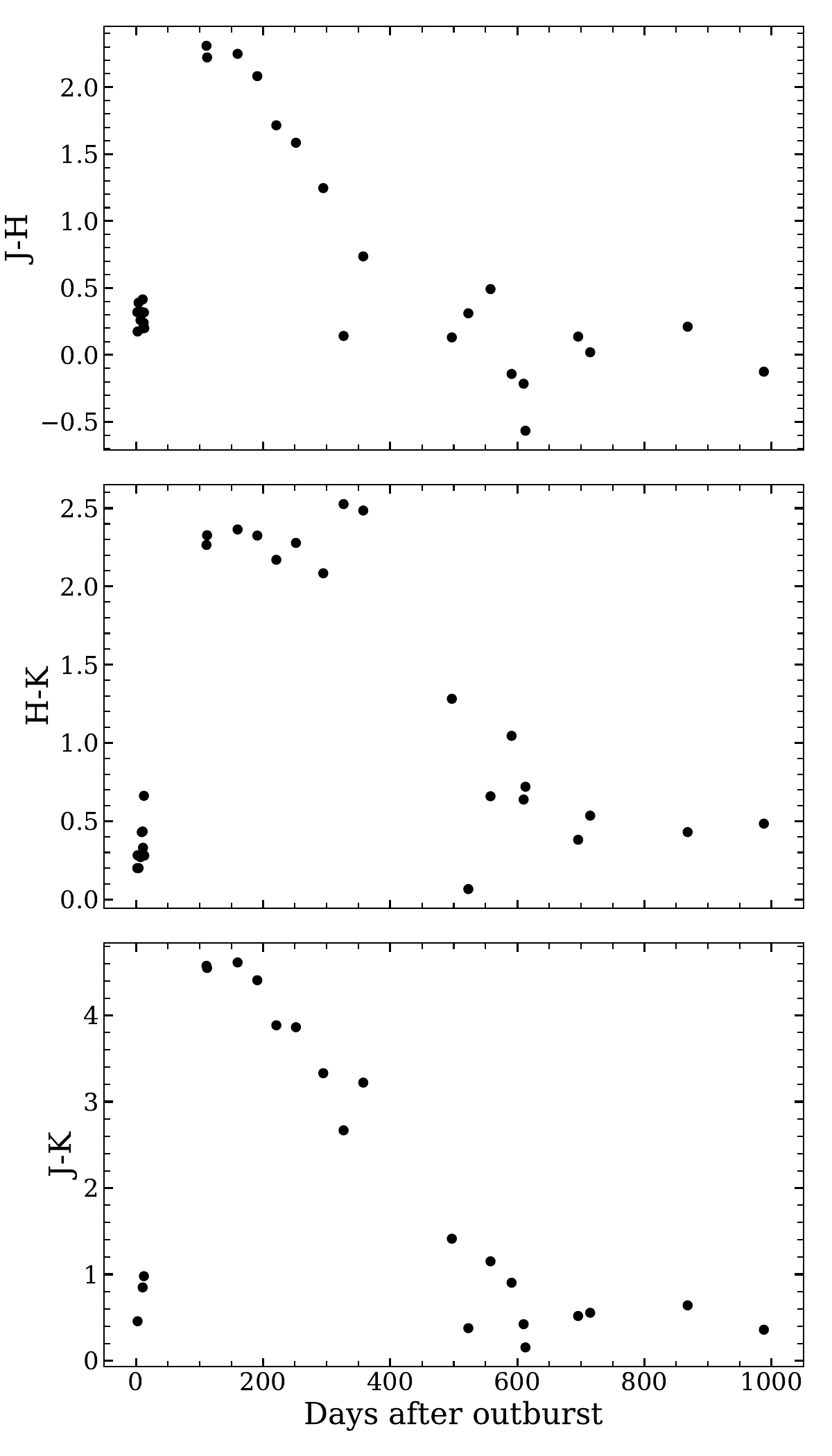}
 \caption{Evolution of NIR color terms of nova V5584 Sgr from day 2 (pre-maximum phase) to 987 (decline phase) since discovery.}
 	\label{nircolorterm}
\end{figure}

The NIR light curves are made using data from \cite{raj2015near} and SMARTS are presented in Fig. \ref{nirlc}. The evolution of the J-H, H-K and J-K color indices is shown in Fig. \ref{nircolorterm}. These indices are clearly influenced by the presence of dust around the system. The J-K, J-H and H-K colors reach a maximum value of 4.6, 2.2 and 2.4 mag, respectively after 159 days. The maximum values of these colors indicate substantial dust formation within the ejecta between 15 and 110 days. 
After this period, the color indices decreased. During the dust condensation phase, longer wavelength fluxes increase due to thermal emission from the dust.
However, once grain growth stops, the flux decreases rapidly as the dust shell expands, thins out,
and becomes less prominent \citep{Gehrz_2008}.
There are no optical and NIR observations between these days due to solar conjunction. Generally during optically thick dust formation, a sharp fall in the $V$ band light curve is observed due to the obscuration of light. However, because of the lack of observations in this phase, we cannot confirm any fall. Although, after 110 days the $B$ and $V$ band magnitudes are in recovery phase (110 to 159 days), indicating that they were likely obscured by dust. Nova Aql 1993 (V1419 Aql) also formed optically thick dust with a similar value of t$_2$ \citep{1996AstL...22..170S}.
There was evidence of dust in the ejecta between 40 to 165 days since discovery, on day 40 optical light curve showed optical dip which recovered around 165 days \citep{strope2010catalog}.
High J-K values are also observed for other dust-forming novae; 
3.79 for V496 Sct \citep{raj2012v496}, 
8.00 for V2676 Oph \citep{raj2017v2676}, and 4.5 for V1831 Aquilae \citep{banerjee2018near}. On the final observation (day 987) of the NIR light curve, V5584 Sgr had magnitudes of 17.4, 17.5, and 17.0 in the J, H, and K bands, respectively.

\subsection{Spectral evolution}
\begin{figure*}
\includegraphics[width=2.2\columnwidth]{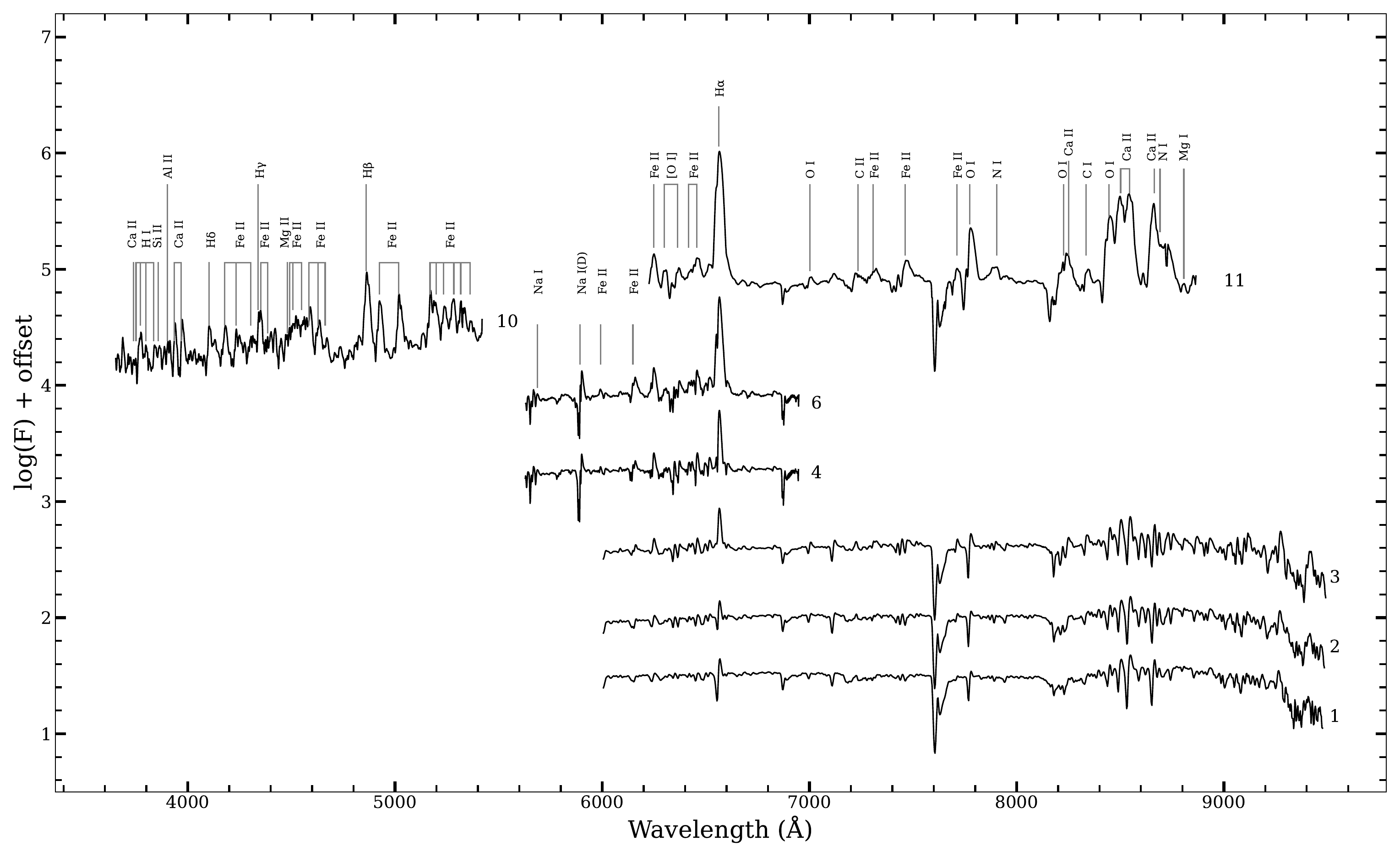}
	\caption{Low-resolution optical spectral evolution of V5584 Sgr obtained from day 1 (Oct 28, 2009) to day 11 (Nov 7, 2009). Spectra are dominated by Fe II multiplets and hydrogen Balmer lines. The lines identified are marked, and time since discovery (in days) is marked against each spectrum.}
	\label{spectra1}
\end{figure*}

\begin{figure*}
\includegraphics[width=2.2\columnwidth]{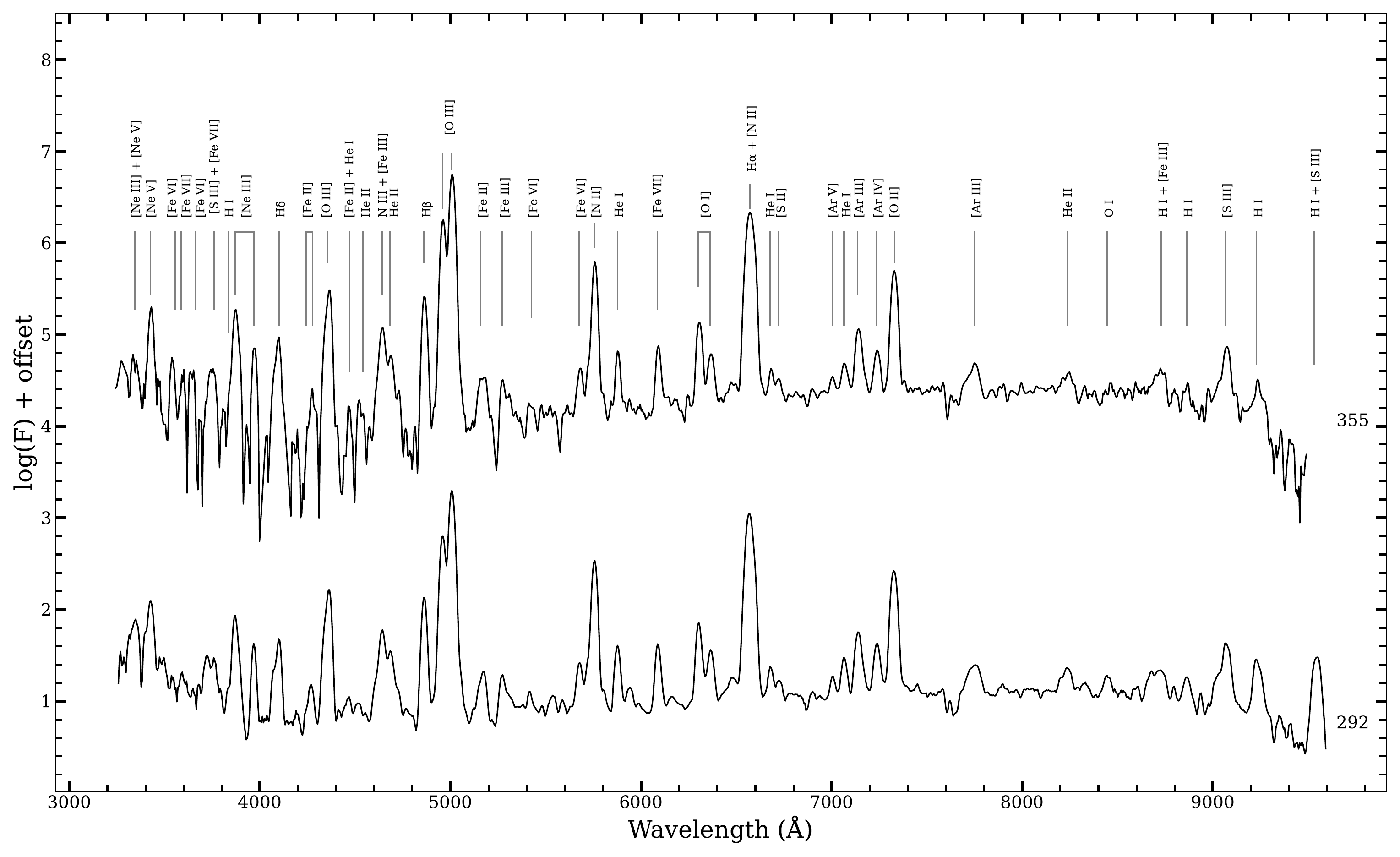}
	\caption{Low-resolution optical spectral evolution of V5584 Sgr obtained on day 292 (Aug 15, 2010) and day 355 (Oct 15, 2010). In Both the nebular phase spectrum [O III] lines at 4363, 4959, 5007 \AA {} and [N II] lines at 5755 \AA{} are very prominent. The lines identified are marked, and time since discovery (in days) is marked against each spectrum.}
	\label{nebular}
\end{figure*}

The optical spectra from SMARTS are presented in Figs. \ref{spectra1} and \ref{nebular}. We presented nine low-dispersion spectra spanning from October 28.02 UT, 2009 to October 17.02 UT, 2010 (day 1 to 355). The first two spectra were taken in the pre-maximum phase, five during early decline, and two in the later nebular phase. The pre-maximum spectra on day 1 and 2 showed very weak emission from H$\alpha$, Fe II 6248, 6456 \AA{} and O I lines 7002, 7773 and 8446 \AA. No other line seen in emission. All of the lines showed deep P-Cygni profiles with a strong absorption component. Other strong absorption include Ca II 8543 and 8662 \AA, C I 7113 \AA {} and Fe II lines.     
The spectra taken after the maximum brightness displayed strong emission lines with weak P-Cygni component from day 3 to 11. The Na II doublet at 5890 and 5896 \AA{} is also seen in our day 4 and 6 spectra. Spectra taken on day 10 and 11 displayed many emission lines including the Balmer lines, [O I] 6300 and 6363 \AA{} and multiple Fe II lines 4629, 4924,  5018, 6248, 6456 \AA. In these epochs, H$\alpha$ was the strongest line followed by the H $\beta$ and Fe II lines. The FWHM velocity was estimated around 1348 $\pm30$ \kms for H$\alpha$ and 1473 $\pm30$ \kms for H$\beta$ line. Following the classification scheme outlined by \cite{williams1992formation}, V5584 Sgr falls within the ``Fe II" class of novae. Novae belonging to this class typically display narrower spectral lines, presence of deep P Cygni absorption profiles, and slower spectral evolution. Additionally, based on the recent universal spectral evolution for novae as described by \cite{aydi2024revisiting}, the spectra during the first 11 days of the eruption were dominated by P Cygni profiles of Balmer and Fe II lines, which remained prominent throughout this period. This indicates that the nova V5584 Sgr was observed during the Fe II phase, near the visible peak and early decline.

Several neutral lines, such as O I (7002, 7773 and 8446 \AA), C I (7115 \AA), N I (7904 and 8692 \AA) and Mg I (8807 \AA), were also present on day 11, likely originating from the low-temperature zone of the nova ejecta where molecule formation can occur. This is supported by the detection of CO band emission in the NIR spectra by \cite{raj2015near} on day 10.2. 
The presence of C I, Mg I, and Na I in the early optical and NIR spectra \citep{raj2015near} clearly indicates a low-temperature zone that could be conducive to dust formation in the nova ejecta. The excess in the NIR colors, J-H = 2.3, H-K = 2.3, and J-K = 4.5, on day 110 also supports dust formation.

By day 292, V5584 Sgr had evolved into the nebular phase of its evolution. Optical spectra taken on June 4 (day 220) and August 10, 2010 (day 287), discussed by \cite{poggiani2011spectroscopic}, also showed that the nova had entered the nebular phase. Our optical spectra (Fig. \ref{nebular}) shows that on day 292 emission was dominated by lines from the hydrogen Balmer series, helium and forbidden emission from metals such as; C, O, N, Ne, Ar, and Fe. The strongest emission lines in the spectra arises from [O III] 4959 and 5007 \AA {} followed by blend of H$\alpha$ with [N II] 6584 \AA,  lines of [N II] 5755, line of [Ne III] 3869 \AA{} and other forbidden lines of oxygen. No major changes were observed in the spectra taken on day 355. However, the line fluxes of the forbidden lines had increased.

\begin{figure}
	\includegraphics[width=1\columnwidth]{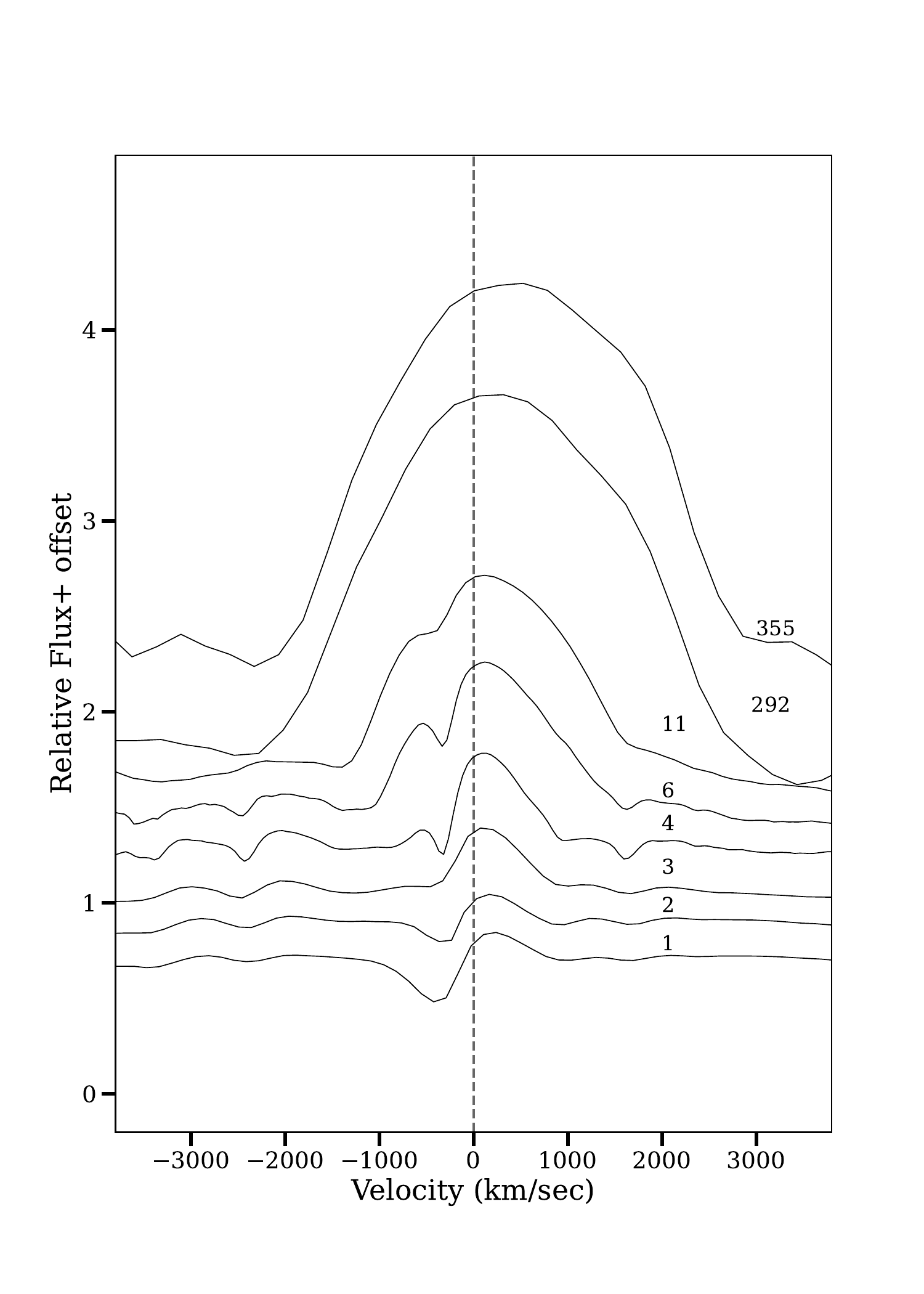}
	\caption{Evolution of H$\alpha$ velocity profile of V5584 Sgr from day 1 (Oct 28, 2009) to 355 (Oct 17, 2010) obtained from spectroscopic data. The line profiles evolve from P-Cygni in the early decline phase to complete emission in the nebular phase.}
	\label{halpha}
\end{figure}

The H$\alpha$ velocity profile evolution is shown in Fig. \ref{halpha}. P-Cygni absorption component of H$\alpha$ at -430 \kms on day 1 changed to -365 \kms on day 2. During the early decline phase, the FWHM velocity ranged from 900 to 1400 \kms. By day 292, it was around 1750 \kms with the blending of the [N II] line.


\begin{table}
\caption{Observed and best-fit NIR \texttt{CLOUDY} model line flux ratios for day 13 of V5584 Sgr}
\label{d13_nir}
\begin{center}
\resizebox{\hsize}{!}{%
\begin{tabular}{lclcc}
\hline
\hline
\textbf{Line ID} & \boldmath{$\lambda$ ($um$)} & \textbf{Observed}$^{a}$ & \textbf{Modelled}$^{a}$ & \boldmath{$\chi^2$}  \\
                    \hline
\  &  & J-Band &  &  \\  
                     \hline
                O I + C I     & 1.1295     & 1.15E$+$00 & 9.38E$-$01 & 5.18E$-$01 \\
                C I           & 1.1409     & 3.53E$-$01 & 5.62E$-$01 & 4.85E$-$01 \\ 
                C I           & 1.1659     & 1.69E$-$01 & 1.09E$-$01 & 5.92E$-$02 \\
                C I           & 1.1755     & 3.69E$-$01 & 3.62E$-$01 & 9.39E$-$04 \\
                C I           & 1.1819     & 4.15E$-$02 & 1.45E$-$01 & 1.72E$-$01 \\
                N I           & 1.2469     & 5.55E$-$02 & 1.40E$-$02 & 2.76E$-$02 \\
                C I           & 1.2562     & 7.15E$-$02 & 7.74E$-$02 & 5.54E$-$04 \\
                Pa $\beta$    & 1.2818     & 1.00E$+$00 & 1.00E$+$00 & 0.00E$+$00 \\
                 O I          & 1.3164    & 1.13E$-$01 & 5.22E$-$02 & 4.16E$-$02 \\
                \hline
\  &  & H-Band &  &  \\
\hline
                CI + Mg I     & 1.5749     & 2.76E$-$01 & 1.50E$-$01 & 1.82E$-$01 \\
                Br 14         & 1.5881     & 5.73E$-$01 & 8.27E$-$01 & 1.03E$+$00 \\
                C I           & 1.6005     & 1.77E$-$01 & 3.09E$-$01 & 4.39E$-$01 \\
                Br 13         & 1.6109     & 3.45E$-$01 & 8.96E$-$01 & 3.37E$+$00 \\
                C I           & 1.6334     & 5.06E$-$01 & 5.07E$-$01 & 3.00E$-$06 \\
                Br 12         & 1.6407     & 1.00E$+$00 & 1.00E$+$00 & 0.00E$+$00 \\
                C I           & 1.6465     & 1.67E$-$01 & 3.11E$-$01 & 3.32E$-$01 \\
                Br 11 + C I   & 1.6865     & 3.91E$+$00 & 3.03E$+$00 & 8.58E$+$00 \\
                C I           & 1.7045     & 1.73E$-$01 & 2.40E$-$01 & 4.91E$-$02 \\
                Mg I  + C I   & 1.7109     & 2.57E$-$01 & 2.05E$-$01 & 4.29E$-$02 \\
                Br 10 + C I   & 1.7390     & 4.18E$+$00 & 4.04E$+$00 & 2.11E$-$01 \\
                C I           & 1.7672     & 1.01E$+$00 & 9.99E$-$01 & 1.30E$-$03 \\
                Blnd C I      & 1.7845     & 1.75E$+$00 & 1.91E$+$00 & 3.09E$-$01 \\
                C I           & 1.7918     & 4.24E$-$01 & 1.02E$+$00 & 3.95E$+$00 \\
                \hline
\  &  & K-Band &  &  \\
\hline
                He I           & 2.0585     & 1.39E$-$01 & 8.21E$-$02 & 5.20E$-$02 \\
                C I            & 2.1023     & 8.57E$-$02 & 4.36E$-$01 & 1.37E$+$00 \\
                Blnd C I       & 2.1259     & 1.86E$-$01 & 2.61E$-$01 & 9.20E$-$02 \\
                Br $\gamma$    & 2.1655     & 1.00E$+$00 & 1.00E$+$00 & 0.00E$+$00 \\
                Na I           & 2.2056     & 2.75E$-$01 & 2.10E$-$01 & 1.07E$-$01 \\
\hline
 \end{tabular}}
\end{center}
\end{table}

\begin{table}
\caption{Best-fit NIR \texttt{CLOUDY} model parameters obtained on day 13 for the system V5584 Sgr}
\label{d13_ph_nir}
\begin{center}
\resizebox{\hsize}{!}{%
\begin{tabular}{l c }
\hline\hline
\textbf{Parameter} & \textbf{Day 13} \\ [0.5ex]
\hline
T$_{BB}$ ($\times$ 10$^{4}$ K) & 1.28 $\pm$ 0.03 \\ [0.25ex]
Luminosity ($\times$ 10$^{36}$ erg/s) & 1.10 $\pm$ 0.04 \\ [0.25ex]
Clump Hydrogen density ($\times$ 10$^{11}$ cm$^{-3}$) & 7.08  \\ [0.25ex]
Diffuse Hydrogen density ($\times$ 10$^{9}$ cm$^{-3}$) & 1.41 \\ [0.25ex]
Covering factor (clump) & 0.70 \\ [0.25ex]
Covering factor (diffuse) & 0.30 \\ [0.25ex]
$\alpha$ & -3.00 \\ [0.25ex]
Inner radius ($\times$ 10$^{13}$ cm) & 7.67 \\ [0.25ex]
Outer radius ($\times$ 10$^{14}$ cm) & 1.53 \\ [0.25ex]
Filling factor  & 0.1 \\ [0.25ex]
He/He$_{\odot}$ & 1.8 $\pm$ 0.4 (1)$^{a}$ \\ [0.25ex]
C/C$_{\odot}$ & 10.8 $\pm$ 3.0 (15) \\ [0.25ex]
N/N$_{\odot}$ & 45.5 $\pm$ 5 (1) \\ [0.25ex]
O/O$_{\odot}$ & 30.2 $\pm$ 4.0 (2) \\ [0.25ex]
Na/Na$_{\odot}$ & 1.55 $\pm$ 0.3 (2) \\ [0.25ex]
Mg/Mg$_{\odot}$ & 1.68 $\pm$ 0.2 (2) \\ [0.25ex]
Ejected mass ($\times$ 10$^{-4}$ M$_{\odot}$) & 1.63 \\ [0.25ex]
Number of observed lines (n) & 28 \\ [0.25ex]
Number of free parameters (n$_{p}$) & 12 \\ [0.25ex]
Degrees of freedom ($\nu$) & 16 \\ [0.25ex]
Total $\chi^{2}$ &  21.45 \\ [0.25ex]
$\chi^{2}_{red}$ & 1.34 \\ [0.25ex]
\hline
\end{tabular}}
\end{center}
$^{a}$The number of lines available to obtain an abundance estimate is as shown in the parenthesis.
\end{table}

\subsection{Physical parameters}
We have used dereddened line fluxes of oxygen lines to estimate the physical parameters of the nova ejecta (e.g. optical depth of oxygen, electron temperature, electron density, and the mass of oxygen). We utilized the spectrum from day 292 (August 15, 2010) to derive the diverse physical parameters of the nova material.

The optical depth of [O I] 6300 \AA\ can be determined using the formula given by \cite{wil94}, 
\begin{align}
\label{op_dep}
\dfrac{F_{\lambda6300}}{F_{\lambda6364}} = \dfrac{(1 - e^{-\tau})}{(1-e^{-\tau/3})}.
\end{align}
The calculated optical depth ($\tau_{6300}$) is 1.36 $\pm 0.15$ on day 292, indicating optically thin ejecta. 
Furthermore, the values of $\tau$ and T$_e$ (5000K typical value from \cite{ederoclite2006} ; \cite{raj2017v2676}) are useful to determine the mass of oxygen in the ejecta using 6300\AA\ line, as given by \cite{wil94}:
 \begin{equation}
\dfrac{m(O)}{M_{\odot}} = 152 d^{2} exp[\dfrac{22850}{T_e}] \times 10^{1.05E(B-V)}\frac{\tau}{1 - e^{-\tau}} F_{\lambda6300}
\end{equation}
We find M$_{O I} = 2.3\times 10^{-6} M_{\odot}$. 
Electron density can be determined by adopting the typical temperature of 5000 K, as assumed above, and [O III] line ratio as in \cite{ost06},
\begin{equation}
\frac{j_{4959}+j_{5007}}{j_{4363}} = 7.9 \frac{e^{3.29\times10^{4}/T_{e}}}{1+4.5\times10^{-4}\frac{N_{e}}{T_{e}^{1/2}}}
\end{equation}

The low-resolution spectra showed a blend of the [O III] 4363 line with H$\gamma$ at 4340 \AA. Using the SPLOT deblending tool in IRAF\footnote[3]{NOIRLab IRAF is distributed by the Community Science and Data Center at NSF NOIRLab, managed by the Association of Universities for Research in Astronomy (AURA) under a cooperative agreement with the U.S. National Science Foundation.}, we resolved the emission lines and calculated the flux for [O III] 4363 \AA. The electron density $N_e$ was computed to be (8.7 $\pm$ 0.55) $\times$ 10$^{7}$ cm$^{-3}$.

\begin{table}
	\caption{Observed and best-fit optical \texttt{CLOUDY} model line flux ratios for day 292 of V5584 Sgr}
	\label{d292_c}
	\begin{center}
		\resizebox{\hsize}{!}{%
			\begin{tabular}{lclcc}
				\hline
				\hline
				\textbf{Line ID} & \boldmath{$\lambda$ (\AA)} & \textbf{Observed}$^{a}$ & \textbf{Modelled}$^{a}$ & \boldmath{$\chi^2$}  \\
				\hline
			    H I           & 3835       & 3.96E$-$02 & 8.07E$-$02 & 2.69E$-$02 \\
				{[}Ne III{]}  & 3869       & 1.42E$+$00 & 1.43E$+$00 & 2.24E$-$03 \\
				{[}Ne III{]}  & 3968       & 5.53E$-$01 & 4.36E$-$01 & 3.40E$-$01 \\
				{[}Si II{]}    & 4069       & 7.26E$-$02 & 7.14E$-$02 & 3.20E$-$05 \\
				H I           & 4102       & 4.08E$-$01 & 2.64E$-$01 & 3.35E$-$01 \\
				{[}Fe II{]}   & 4244       & 3.59E$-$02 & 9.21E$-$02 & 5.64E$-$02 \\
				{[}Fe II{]}   & 4276       & 1.04E$-$01 & 5.81E$-$02 & 2.36E$-$02 \\
                H I + {[}O III{]}   & 4363       & 2.32E$+$00 & 2.73E$+$00 & 1.89E$+$00 \\
				  N III + {[}Fe III{]}       & 4645       & 3.64E$-$01 & 2.33E$-$01 & 1.88E$-$01 \\
				He II         & 4686       & 1.10E$-$01 & 3.89E$-$01 & 1.24E$+$00 \\
				H I           & 4861       & 1.00E$+$00 & 1.00E$+$00 & 0.00E$+$00 \\
				  {[}O III{]}   & 4959       & 3.15E$+$00 & 3.78E$+$00 & 4.52E$+$00 \\
				  {[}O III{]}   & 5007       & 12.09E$+$00 & 11.30E$+$00 & 7.01E$+$00 \\
				  {[}Fe II{]}   & 5159       & 1.31E$-$01 & 1.74E$-$01 & 2.10E$-$02 \\
                {[}Fe III{]}  & 5270       & 7.78E$-$02 & 1.50E$-$01 & 8.35E$-$02 \\
				{[}Fe VI{]}   & 5424       & 1.85E$-$02 & 1.41E$-$02 & 2.26E$-$04 \\
				{[}Fe VI{]}   & 5677       & 7.30E$-$02 & 2.69E$-$02 & 3.39E$-$02 \\
				{[}N II{]}    & 5755       & 1.62E$+$00 & 1.66E$+$00 & 3.16E$-$02 \\
				He I          & 5876       & 1.72E$-$01 & 8.78E$-$02 & 7.95E$-$02 \\
				{[}Fe VII{]}  & 6086       & 1.65E$-$01 & 5.18E$-$02 & 1.42E$-$01 \\
				{[}O I{]}     & 6300       & 2.71E$-$01 & 7.03E$-$01 & 2.77E$+$00 \\
				{[}O I{]}     & 6364       & 1.32E$-$01 & 2.24E$-$01 & 1.35E$-$01 \\
				H I + {[}N II{]} & 6572    & 5.38E$+$00 & 4.00E$+$00 & 21.42E$+$00 \\
				{[}Ar V{]}    & 7006       & 1.72E$-$02 & 9.82E$-$03 & 8.81E$-$04 \\
                He I          & 7065       & 4.00E$-$02 & 6.26E$-$03 & 1.26E$-$02 \\
				{[}Ar III{]}  & 7136       & 1.78E$-$01 & 2.18E$-$01 & 3.98E$-$02 \\
                {[}O II{]}    & 7320        & 8.83E$-$01 & 6.38E$-$01 & 6.67E$-$01 \\
                {[}Ar III{]}  & 7751       & 8.30E$-$02 & 5.18E$-$02 & 1.07E$-$02 \\
                 He I         & 8237       & 3.61E$-$02 & 8.21E$-$03 & 8.68E$-$03 \\
                 H I          & 8863       & 1.92E$-$02 & 1.47E$-$02 & 5.07E$-$04 \\
                 {[}S III{]}  & 9069       & 9.30E$-$02 & 1.01E$-$01 & 1.64E$-$03 \\
                 H I          & 9229       & 5.24E$-$02 & 2.50E$-$02 & 1.88E$-$02 \\
                  {[}S III{]}  & 9531       & 7.60E$-$02 & 2.24E$-$01 & 5.07E$-$01 \\
				\hline
		    \end{tabular}}
	\end{center}
	$^{a}$Relative to H$\beta$
\end{table}
\begin{table}
	\caption{Best-fit optical \texttt{CLOUDY} model parameters obtained on day 292 for the system V5584 Sgr}
	\label{d292_phy_opt}
	\begin{center}
		\resizebox{\hsize}{!}{%
			\begin{tabular}{l c } 
				\hline\hline
				\textbf{Parameter} & \textbf{Day 292} \\ [0.5ex] 
				\hline 
				T$_{BB}$ ($\times$ 10$^{5}$ K) & 2.82 $\pm$ 0.10 \\ [0.25ex]
				Luminosity ($\times$ 10$^{36}$ erg/s) & 3.06 $\pm$ 0.50 \\ [0.25ex]
				Clump Hydrogen density ($\times$ 10$^{7}$ cm$^{-3}$) & 1.27  \\ [0.25ex]
				Diffuse Hydrogen density ($\times$ 10$^{6}$ cm$^{-3}$) & 4.54 \\ [0.25ex]
				Covering factor (clump) & 0.60 \\ [0.25ex]
				Covering factor (diffuse) & 0.40 \\ [0.25ex]
				$\alpha$ & -3.00 \\ [0.25ex]
				Inner radius ($\times$ 10$^{15}$ cm) & 2.82 \\ [0.25ex]
				Outer radius ($\times$ 10$^{15}$ cm) & 5.66 \\ [0.25ex]
				Filling factor  & 0.10 \\ [0.25ex]
				N/N$_{\odot}$ & 24.48 $\pm$ 3.0 (2)$^{a}$ \\ [0.25ex]
				O/O$_{\odot}$ & 4.50 $\pm$ 0.8 (5) \\ [0.25ex]
				Fe/Fe$_{\odot}$ & 1.20 $\pm$ 0.10 (8) \\ [0.25ex]
				Ne/Ne$_{\odot}$ & 1.50 $\pm$ 0.10 (2) \\ [0.25ex]
				Ejected mass ($\times$ 10$^{-4}$ M$_{\odot}$) & 1.55 \\ [0.25ex]
				Number of observed lines (n) & 33 \\ [0.25ex]
				Number of free parameters (n$_{p}$) & 10 \\ [0.25ex]
				Degrees of freedom ($\nu$) & 23 \\ [0.25ex]
				Total $\chi^{2}$ & 41.65 \\ [0.25ex]
				$\chi^{2}_{red}$ & 1.81 \\ [0.25ex]
				\hline
		\end{tabular}}
	\end{center}
	$^{a}$The number of lines availed to obtain abundance estimate is as shown in the parenthesis.
\end{table}

\begin{figure*}
\includegraphics[width=2.1\columnwidth]{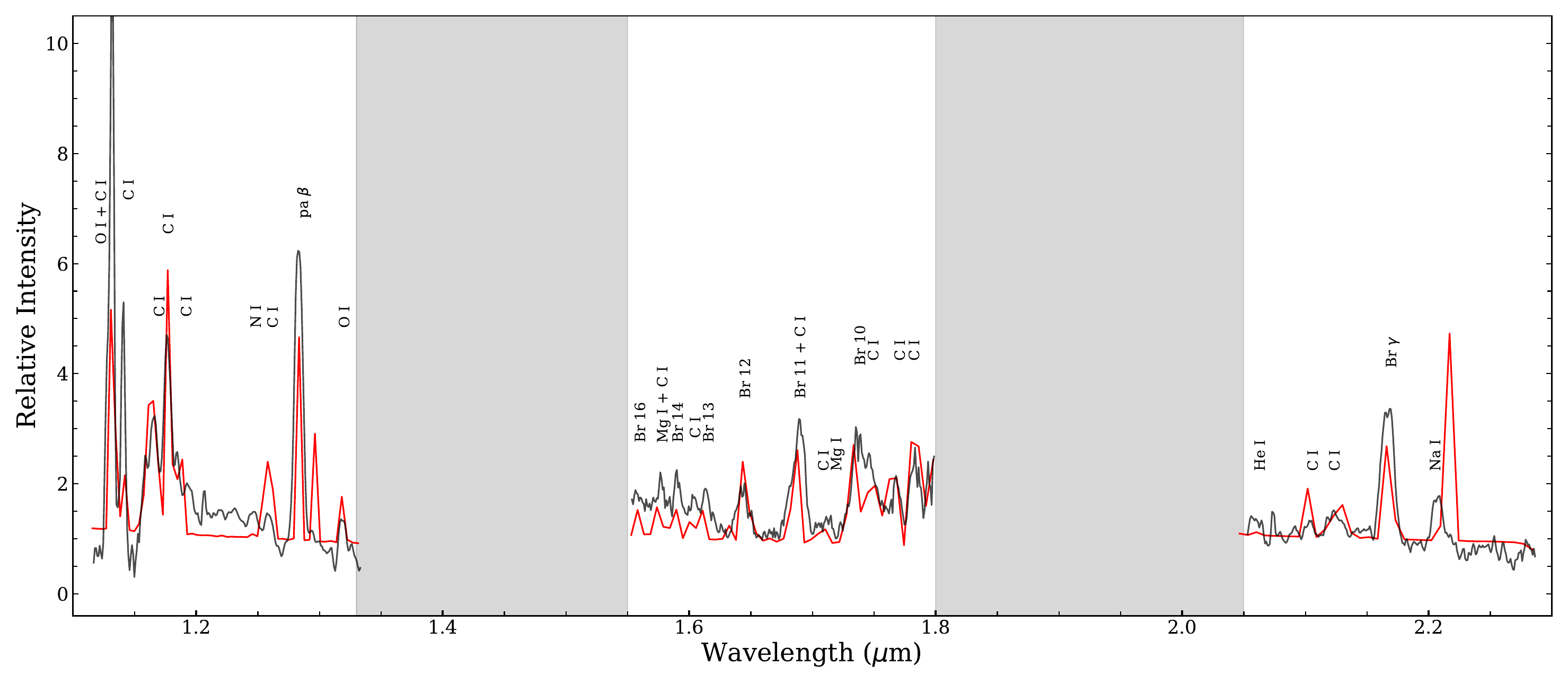}
	\caption{Best-fit NIR JHK band \texttt{CLOUDY} synthetic spectrum (red line) plotted over the observed spectrum (black line) of V5584 Sgr obtained on Nov 08, 2009 (day 13).}
	\label{d13_1d}
\end{figure*}

\begin{figure*}
	\includegraphics[scale=0.47]{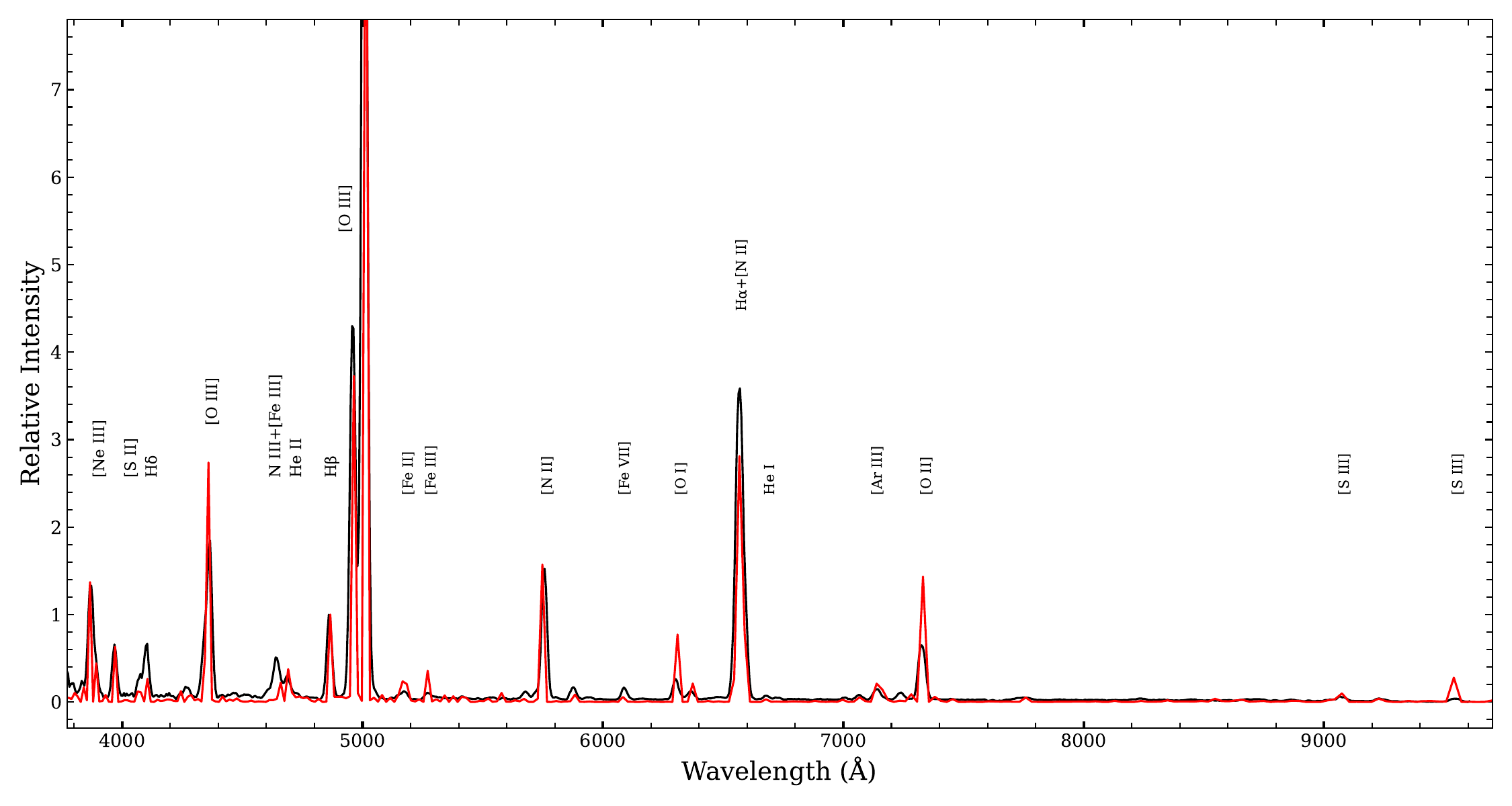}
	\caption{Best-fit optical \texttt{CLOUDY} synthetic spectrum (red line) plotted over the observed spectrum (black line) of V5584 Sgr obtained on Aug 15, 2010 (day 292).}
	\label{d292_1d}
\end{figure*}

\subsection{Photoionization Modeling of optical and NIR spectra}
The photoionization spectral synthesis code \texttt{CLOUDY}
(version C23.01 \citet{chatzikos2023}) was used to
understand the physical conditions and elemental abundances of V5584 Sgr. In the present analysis, we modeled the day 13 NIR JHK spectra from \cite{raj2015near} and the day 292 optical spectra. By selecting the spectra from the ``pre-dust" phase (13 days) and the ``nebular phase" (292 days), we focused on epochs with minimal dust interference, allowing for a more accurate analysis of elemental emissions. \texttt{CLOUDY} solves the equations of thermal and statistical equilibrium using the input physical parameters of the system to compute the intensities and column densities of numerous spectral lines ($\sim$$10^4$) across the electromagnetic spectrum. The predicted fluxes of these emission lines are then directly compared with observed fluxes to estimate the physical parameters of the source and the chemical abundances of the gas cloud. In its default mode, \texttt{CLOUDY} accounts for photoionization and collisional
ionization. In the current model, both processes are active, making it not strictly a pure
photoionization model. However, most lines are produced via photoionization and
recombination \citep{woodward2024photoionization}.
To model the spectra of V5584 Sgr, we assumed a central ionising radiation having blackbody shape with temperature T$_{BB}$(K) and luminosity L (erg sec$^{-1}$), surrounded by a spherically symmetric expanding shell. The shell parameters include the density, inner and outer radii, geometry, cover factor, filling factor, and elemental abundances within the shell. 
The density of the shell is determined by the total hydrogen number density, n(r) (cm $^{-3}$). 
We assume the surrounding shell is clumpy and represent its clumpiness using the filling factor parameter. In our model, the radial variations of the hydrogen density, n(r), and the filling factor, f(r), are defined as follows:
\begin{equation}
    n(r) = n(r_{in})(r/r_{in})^{\alpha}
\end{equation}
\begin{equation}
    f(r) = f(r_{in})(r/r_{in})^{\beta}
\end{equation}
where r$_{in}$ is the inner radius, $\alpha$ and $\beta$ are exponents of the power laws. Both the exponents are typical of those used in other novae in similar photoionization studies \citep[e.g.,][and references therein]{2010AJHelton,raj2018cloudy2676,2020MNRASPavana,raj2024dustyaftermathrapidnova} and are held at these values to reduced the number of free parameters. For the present study, we have used typical value of the power-law index \(\alpha = -3\) assuming a constant mass per unit volume throughout the ejecta. The filling factor in novae ejecta typically ranges from 0.01 to 0.1 \citep{Shore_2008}. During the early stage the value of filling factor generally is around 0.1, which decreases with time \citep{ederoclite2006}. 
Our model employs a filling factor of 0.1 and a power-law index \(\beta = 0\). The inner and outer radii of the model shell are derived from the observed minimum and maximum expansion velocities,
calculated from the FWHM of emission lines and the time since the outburst.
Based on previously studied novae, we assigned ranges to all free parameters, including blackbody temperature, luminosity, density, and elemental abundances for elements with emission lines present in the observed spectra. The abundances of other elements were kept at solar values as given by \citet{2010Ap&SSGrevesse}. We have varied all the above free parameters in the small steps simultaneously. Initially, our focus was on generating prominent lines such as hydrogen and carbon in the early NIR spectra, and oxygen and argon in the nebular optical spectra. Subsequently, we varied the elemental abundances to match the flux of these lines with the observed spectra. A similar approach was used for matching fainter lines. The initial selection of the best models relied solely on visual inspection. Many spectra were generated to get the best fit for both epochs. However, the final best-fit model was determined by assessing the goodness of fit by calculating the values $\chi^{2}$ and reduced $\chi^{2}$. These parameters were obtained using the following relation:

\begin{equation}
    \chi^{2} = \sum_{i=1}^{n} \dfrac{(M_{i} - O_{i})^{2}} {\sigma^{2}_{i}}\\
\end{equation}

\begin{equation}
    \chi^{2}_{red} = \dfrac{\chi^{2}}{\nu}
\end{equation}
M$_{i}$ and O$_{i}$ define the modeled and observed line ratios. The symbol $\sigma_{i}$ shows the error in the observed flux ratio. The degrees of freedom, represented by $\nu$, are calculated as \( n - n_{p} \), where \( n \) is the number of observed lines and \( n_{p} \) is the
number of free parameters.
 
Typically, the error ($\sigma$) falls within a range of 10\% to 30\%. This range depends on several factors, including the strength of the spectral line relative to the continuum, the possibility of blending with other spectral lines, uncertainties in de-reddening value, and errors in the measured line flux \citep{2010AJHelton,woodward2024photoionization}. We considered $\sigma$ = 20\% to 30\%  for the present study. For an appropriate fit, the values of $\chi^{2}$ $\sim$ $\nu$ and $\chi^{2}_{red}$ should be low, usually between 1 and 2. The observed emission lines are dereddened using \textit{E(B-V)} value obtained in section \ref{reddening and distance}.
We measured the fluxes of each line by fitting Gaussian profiles using the \texttt{splot} task in IRAF. To reduce uncertainties associated with flux calibration
across different wavelength regions, we calculated the ratio of modeled and observed fluxes
relative to strong, unblended hydrogen lines, namely H$\beta$ in the optical region, Pa $\beta$ in the $J$ band, Br 12 in the $H$ band, and Br $\gamma$ in the $K$ band.

Initially we started the modeling with one density component which reproduced the majority of prominent hydrogen and low-ionization emission lines but failed to generate the higher ionization lines. Nova ejecta typically exhibit an inhomogeneous density structure, with clumps and regions of varying density observed within the expanding material \citep{paresce1995structure,BodeEvansBook2008,williams2013density,2021ARA&A..59..391C}. To address this, an additional low-density component was added in order to match these lines. As a result of the lower density, the ionizing photons penetrate more deeply into the shell, resulting in a hotter and more ionized shell \citep{shore2003early}. 
Both components shared the same physical parameters, except for the hydrogen density and covering factor, which increases the number of free parameters by two. The covering factor represents a fraction of 4$\pi$ sr enclosed by the model shell. We set the covering factors for both components so that their sum is less than or equal to 1. After
multiplying the spectrum of the high-density (clump) and low-density (diffuse) components by their respective covering factors, the resulting spectrum was obtained by adding the spectra from both components.

We used 28 emission lines in the early NIR spectra and 33 lines in the optical spectra for our fit. 
Tables \ref{d13_nir} and \ref{d292_c} present the dereddened relative fluxes of the observed and best-fit model lines, along with the corresponding $\chi^{2}$ values for the NIR and optical spectra, respectively. In both tables, the high values of $\chi^{2}$ are possibly attributed to three main reasons. Firstly, the blending of lines contributes to this value. Secondly, the model's inability to generate certain lines is linked to the consideration of the spherical shape of the ejecta, which is not ideal for novae. Lastly, the inhomogeneous density structure of the ejecta, with a wide range of densities \citep{williams2013density}. However, the reduced $\chi^{2}$ values are in an acceptable range for the appropriate fit. The resulting values for the physical parameters and abundances are given in table \ref{d13_ph_nir} and \ref{d292_phy_opt} for days 13 and 292 respectively. All parameters are in cgs units, and all abundances are given relative to solar values. 
The parentheses in the tables indicate the number of spectral lines used to estimate the abundance. A higher number of lines for an element leads to a more accurate abundance value. Thus, the abundances of nitrogen and oxygen are much more uncertain in the modeling on day 13. 

The estimated abundance values indicate that nitrogen, carbon, and oxygen are more than solar in the early phase, higher carbon and oxygen abundance supports dust formation in V5584 Sgr. Similarly, \cite{raj2024dustyaftermathrapidnova} also estimated high carbon and oxygen abundances in the pre-dust phase of nova V5579 Sgr, which subsequently decreased due to dust grain formation. Our abundance analysis supports the enhanced CNO for CO novae \citep{Shore_2008}. In the nebular phase neon and iron abundance are near to solar values. 
In both epochs, we found that the model demonstrates a nearly constant luminosity, which supports the prediction that novae radiate at a nearly constant luminosity until the nuclear fuel is exhausted on the WD surface \citep{1972ApJ...176..169S}.
Figs. \ref{d13_1d} and \ref{d292_1d} display the best-fit modeled spectrum (red line) along with the corresponding dereddened observed spectrum (black line) for the JHK and optical regions, respectively. On day 13, our model predicts low blackbody temperature and luminosity, which favors the appearance of neutral metal lines in the optical spectra.

The mass of the ejecta within the modeled shell was calculated using the equation provided by \citet{schwarz2001massejecta}:
\begin{equation}
	M_{\text{shell}} = n(r_\text{in}) f(r_\text{in}) \int_{R_{\text{in}}}^{R_{\text{out}}} \left(\frac{r}{r_\text{in}}\right)^{\alpha+\beta} 4\upi r^2 dr
\end{equation}
The parameters for density, filling factor, $\alpha$, and $\beta$ were derived from the best-fit model. The total ejected shell mass was estimated by calculating the mass for each density component (clump and diffuse), multiplying these values by their respective covering factors,
and summing the results.
The mass of the ejected hydrogen shell was estimated to be 1.63 $\times$ 10$^{-4}$ M$_{\odot}$ and 1.55 $\times$ 10$^{-4}$ M$_{\odot}$ based on the modeling results from day 13 and day 292, respectively. 
These values are consistent with the theoretical values estimated for CNe \citep{gehrz1998nucleosynthesis}.

\subsection{Dust mass and temperature} 
\cite{2010russel_atel} reported dust formation within the nova ejecta based on their NIR observations obtained on February 10, 2010 (106 days since discovery).  Additionally, high values of J-H, H-K, and J-K color plots (fig. \ref{nircolorterm}) clearly indicate substantial dust formation. The observations from the Wide field Infrared Survey Explorer (WISE; \citet{wright2010wide}) in 3.4
(W1), 4.6 (W2), 12 (W3) and 22 $\mu$m (W4) bands also support dust formation (see Fig. 9; \citealt{raj2015near}). The onset of dust formation is not known due to the limited coverage of the photometric data resulting from the solar conjunction. %
We expect the onset of dust formation to occur at approximately $45 \pm 5$ days from the maximum, taking t$_2$ = 26 days (see Fig. 2; \citealt{williams2013tdt2}).

To estimate the outburst luminosity we have used the ($\lambda F_{\lambda})_{max}$ values for optical and IR from \cite{raj2015near}. Using the relation given by \citet{Gehrz_2008} that the outburst luminosity is equal to 4.11 $\times$ 10$^{17}$d$^{2}$ ($\lambda F_{\lambda})_{max}$ L$_{\odot}$. We calculate the outburst luminosity (L$_{O}$) and infrared dust luminosity (L$_{IR}$) to be 1.48$\times$10$^{5}$ and 6.71$\times$10$^{3}$L$_{\odot}$, respectively using distance d = 7.4 kpc. Furthermore, the estimated visible optical depth ($\tau = \frac{L_{\text{IR}}}{L_{\text{O}}}$) is 0.05, indicating the existence of optically thin dust emission during the early stages of dust formation. Similar values were also determined for V1668 Cyg \citep{gehrz1988infrared}, V1831 Aql \citep{banerjee2018near}, and V5579 Sgr \citep{raj2024dustyaftermathrapidnova}. The small $\tau$ value may indicate a homogeneous dust shell with insufficient material to achieve optical thickness.
\begin{figure}
    \centering
     \includegraphics[width=1.0\columnwidth]{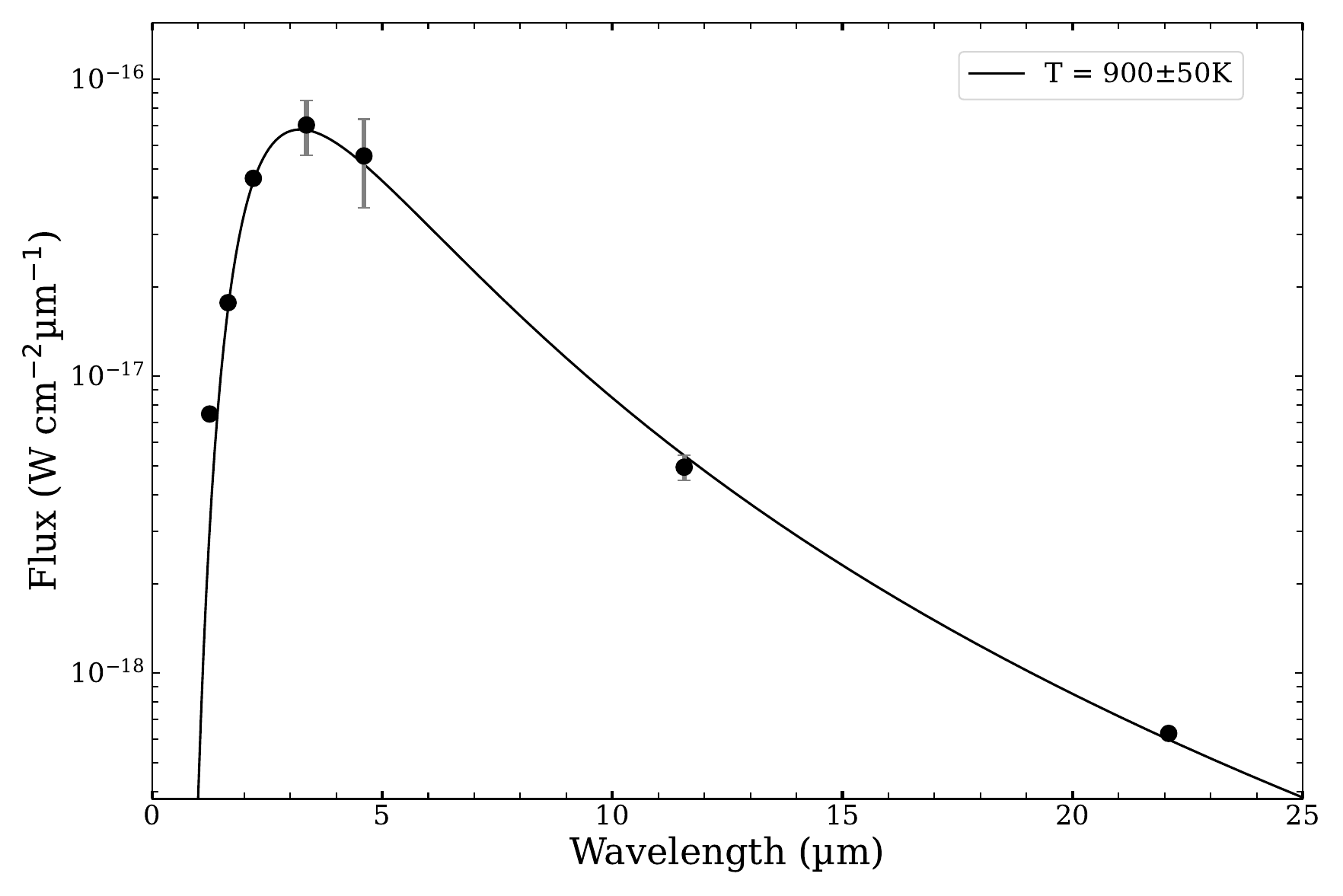}
     \caption{The SED shows a best fit blackbody to the JHK and WISE data taken on day 152, with a temperature of about 900 K.}
    \label{sed_temp}
\end{figure}

This study utilized spectra from day 13 and day 292 for photoionization modeling. Our best-fit \texttt{CLOUDY} models estimate the carbon abundance (C/H) to be 10.8 times the solar value on day 13, decreasing to the solar value by day 292. Conversely, the oxygen abundance (O/H) was 30.2 times the solar value on day 13 and reduced to 4.5 times the solar value at the later stage, indicating that the ejecta remains oxygen-rich. The observed decrease in carbon abundance is likely due to the depletion caused by dust grain formation.
Theoretical hydrodynamic models of the thermonuclear runaway (TNR) also suggest that the gas phase of the nova ejecta has a relatively higher abundance of oxygen. This is because an environment with a higher ratio of oxygen (O) to carbon (C) ratio can facilitate the formation of carbon-rich dust grains \citep{starrfield2016thermonuclear}.

We estimated, using the magnitudes from SMARTS and WISE, the temperature of the dust shell to be 900 $\pm$ 50 K at 152 days post-discovery. We have corrected the JHK magnitudes for reddening before the SED fit using A$_J$=0.70, A$_H$=0.44, and A$_K$=0.28 which were estimated using the ratios A$_J$/A$_V$=0.282, A$_H$/A$_V$=0.175, and A$_K$/A$_V$=0.112 from \cite{rieke1985interstellar}. The SED has peak emission near W1 (3.4 \micron) in fig. \ref{sed_temp}. 
We estimate the dust mass following \cite{evans2017rise} and
\citet{banerjee2018near}, assuming that the grains are spherical and
that the dust is composed of carbonaceous material. 
Using the relations given by \cite{evans2017rise}, we find that the dust masses for the grains of amorphous carbon (AC) and graphitic carbon (GR) are as follows.
For optically thin  amorphous carbon,
\begin{equation}
\frac{M_{dust-AC}}{M_{\odot}} \simeq 1.02 \times 10^{18} \frac{(\lambda f_{\lambda})_{max}}{T^{4.754}_{dust}}
\end{equation}
and for optically thin graphitic carbon,
\begin{equation}
\frac{M_{dust-GR}}{M_{\odot}} \simeq 9.07 \times 10^{19} \frac{(\lambda f_{\lambda})_{max}}{T^{5.315}_{dust}}
\end{equation} 
where a value of $D$ = 7.4 kpc has been used for the distance,  $\rho$ = 2.25 gm cm$^{-3}$ has been taken for  the density of  the carbon grains, and ($\lambda F_{\lambda})_{max}$ is in unit of W/m$^{2}$. The dust mass, which is independent of grain size
\citep{evans2017rise}, is estimated to be $\sim$ 2.24$\times 10^{-8}$ and
4.1$\times 10^{-8}M_{\odot}$ , for AC and GR grains, respectively.
The estimated M$_{wd}$ and low ejecta velocity suggest that V5584 Sgr is a CO nova, which generally produces more dust due to the high ejecta mass\citep{gehrz1998nucleosynthesis}.

\section{Discussion}\label{Discussion}

V5584 Sgr is a moderately fast nova (t$_2$ $\sim$ 26
days) with initial CO detection around 10 days, and the excess values of the NIR colors showing dust had formed by 110 days. The time of onset is not known due to the limited photometric coverage caused by solar conjunction. We expect that the dust formation started around 45 $\pm$ 5 days. Generally, the dust formation time for moderately fast novae ranges from 35 to 105 days. Other moderately fast novae with CO detection are NQ Vul \citep{ferland1979carbon}, V705 Cas \citep{evans1996infrared}, V842 Cen \citep{1991isrs.conf..353W}, V2274 Cyg \citep{rudy2003near}, V2615 Oph \citep{das2009detection}, V2676 Oph \citep{2012CBET.3103....1R}, and V496 Sct \citep{raj2012v496}. All of them have formed dust between 40 and 92 days after the outburst.

Our multiwavelength abundance analysis suggests that the carbon abundance (C/H) was 10.8 times solar value on day 13 (pre-dust), with no lines detected in the subsequent phase (post-dust), indicating that the ejecta might be rich in carbon grain. The decrease in the abundance of carbon can be attributed to its depletion into dust grain formation. In the case of V5579 Sgr, where the carbon abundance also decreases in the post-dust phase, PAHs were found in their mid-NIR spectra \citep{raj2024dustyaftermathrapidnova}. However, it has been observed that some novae can produce significant amounts of carbon-rich and oxygen-rich dust grains. Examples of such cases include V705 Cas \citep{evans2005infrared}, V1065 Cen \citep{2010AJHelton}, and V5668 Sgr \citep{gehrz2018temporal}. The oxygen abundance (O/H), which was 30.2 times the solar value on day 13 (pre-dust) and 4.5 times the solar value on day 292 (post-dust), suggests the likelihood of oxygen-rich dust grain formation in addition to carbon-rich dust.

Our \texttt{CLOUDY} modeling supports the notion of nitrogen enrichment in the ejecta of V5584 Sgr, likely resulting from proton capture processes during the TNR. This result is consistent with findings from other novae, such as V705 Cas, and V5579 Sgr, where \citet{hauschildt1994early}, and \cite{raj2024dustyaftermathrapidnova} observed an enrichment in heavy elements especially carbon, nitrogen, and oxygen relative to solar abundances. \citet{shore2018spectroscopic} investigated the potential effects of a significantly enhanced nitrogen-to-carbon (N/C) ratio, particularly its role in promoting dust formation in CO-type novae. The overall enrichment of heavy elements in the ejecta may disturb the thermal balance of the gas, creating conditions conducive to the formation of dust nucleation sites. As \citet{ferland1978heavy} suggested, higher metallicity enhances the cooling efficiency of the ejecta, allowing thermal equilibrium to be reached at lower kinetic temperatures.

\section{Summary}\label{summary}
The evolution of the optical spectrum of nova V5584 Sgr is presented here based on the data obtained from day 1--355 post-discovery.
We have modeled the early phase NIR spectra and nebular phase optical spectra to estimate physical parameters and chemical abundances. Additionally V5584 Sgr formed dust, so we have discussed dust parameters (temperature and mass). 
The key findings of the analyses are summarized below.
\begin{enumerate}
 \item  Based on optical and NIR data from AAVSO and SMARTS, t$_2$ and t$_3$ were estimated to be 26 $\pm$ 1 d and 48 $\pm$ 2 days, respectively, indicating that V5584 Sgr belongs to the class of moderately fast novae. 
 The reddening, \textit{E(B-V)}, was estimated to be approximately 0.80, with the distance to the nova determined as 7.4 $\pm$ 0.5 kpc. The mass of the white dwarf (WD) was calculated to be 0.92 $\pm$ 0.07 M$_\odot$.
  \item We presented the spectral evolution of V5584 Sgr from day 1 to day 355, covering the pre-maximum, early decline, and nebular phases. The optical spectral line profiles evolved from P-Cygni with broad emission and narrow absorption components to emission.  
  The prominent Balmer and Fe II lines clearly indicated that the nova was observed during the Fe II phase, near the visible peak and early decline.
 \item The photo-ionization analysis was carried out for two epochs of multiwavelength spectra day 13 and 292. We found that the model demonstrates a nearly constant luminosity, for both epochs with average values of (2.08 $\pm$ 0.10) $\times$ 10$^{36}$ erg s$^{-1}$ and an average ejected mass of (1.59 $\pm$ 0.04) $\times$ 10$^{-4}$M$_{\odot}$.
\item Our abundance analysis shows that the ejecta is significantly enhanced relative to solar, O/H = 30.2, C/H = 10.8, He/H = 1.8,  Mg/H = 1.68, Na/H = 1.55, and N/H = 45.5 in the early decline phase and O/H = 4.5, Ne/H = 1.5, and N/H = 24.5 in the nebular phase.
 \item  The observed NIR excess indicates the formation of a significant amount of
optically thick dust within the nova ejecta. After reaching a peak around day 159, the J-K, J-H,
and H-K color indices began to decrease, indicating either the destruction of dust molecules by
radiation from the central source or the thinning of the dust shell as it expanded and became
less prominent. Based on WISE and SMARTS NIR magnitudes, the dust temperature was
determined to be approximately 900 $\pm$ 50 K on day 152. The estimated dust mass is 2.24
$\times 10^{-8}$ M${\odot}$ and 4.1 $\times 10^{-8}$ M${\odot}$ for AC and GR grains, respectively.

Overall, the findings of this study contribute to a more comprehensive understanding of dust formation and chemical enrichment in nova ejecta. The results highlight the complex interplay between the physical conditions in the ejecta and the processes that govern the formation of the molecule and dust. Future observations and modeling efforts will continue to refine our understanding of these dynamic systems.
In conclusion, the multi-wavelength observations and modeling of novae contribute to our understanding of the physical processes driving nova eruptions, particularly in the context of dust formation and chemical enrichment. Further studies, including high-resolution spectroscopy and continued monitoring, will be essential to refine our understanding of these phenomena and the role of novae in the broader context of galactic chemical evolution.

\end{enumerate}

\section*{Acknowledgements}
The authors would like to thank the referee for critically reading the manuscript and providing valuable suggestions to improve it. 
We acknowledge with thanks the variable star observations from the AAVSO International Database contributed by observers worldwide and used in this research. We also acknowledge the use of SMARTS data. G Shaw acknowledges support from DST/WOS-A/PM-2/2021.

\textit{Software:} IRAF (v2.16.1 \cite{tod93}), Python (v3.6.8), \texttt{CLOUDY} (C23.01))\\
\textit{Python modules:}  numpy (v1.17.0 \cite{oli06}), scipy (v1.3.1 \cite{Eri01}), matplotlib (v3.1.1 \cite{Hun07}).  

\section*{Data Availability}
Modeling in this paper uses the \texttt{CLOUDY} code (version c23.01), which is freely available at \url{https://www.nublado.org/}. The observed optical data are available at \url{http://www.astro.sunysb.edu/fwalter/SMARTS/NovaAtlas/}. The NIR data underlying this article will be shared on reasonable request to the corresponding author.

\bibliographystyle{mnras}
\setlength{\bibsep}{0.0pt}
\footnotesize{\bibliography{v5584}}

\label{lastpage}
\end{document}